\documentstyle[12pt]{article}

\def\be{\begin{equation}}
\def\ee{\end{equation}}
\def\bea{\begin{eqnarray}}
\def\eea{\end{eqnarray}}

\author{Gary R. Goldstein\thanks{Supported in part by funds  
provided by the U.S. Department of Energy (D.O.E.) 
\#DE-FG02-92ER40702.}
\thanks{gary.goldstein@tufts.edu}\\
Department of Physics and Astronomy\\
Tufts University\\
Medford, MA 02155  USA\\
and\\
Kameshwar C. Wali\thanks{Supported in part by funds  
provided by the U.S. Department of Energy (D.O.E.) 
\#DE-FG02-85ER402.}
\thanks{wali@physics.syr.edu}\\
Department of Physics\\
Syracuse University\\
Syracuse, NY 13244-1130 USA}

\title{Baryons and Mesons with Beauty}
\begin{document}
\maketitle

\pagebreak
\begin{abstract}
Recent experimental findings of several mesons and baryons with {\it beauty} 
and {\it charm} as flavors remind us of the days when strangeness was discovered, and how its inclusion led to SU(3)-flavor symmetry with enormous success in the classification of the ``proliferatedÓ states into SU(3) multiplets. One of the key elements was the successful application of the first order perturbation in symmetry breaking, albeit what then appeared to be huge mass differences, and the prediction of new states that were confirmed by experiments. In this note, we venture into the past and, applying the same techniques, predict some new {\it beauty}- and {\it charm}- flavored hadrons. If these new states are confirmed experimentally, it may provide a useful phenomenological model for classifying numerous states that are found to be in the PDG data and could invite further theoretical challenges towards our understanding of symmetry breaking. 
\end{abstract}

\pagebreak

It is well known that Gell-Mann's eightfold way~\cite{g-mn} leading to the
classification of the then known mesons and baryons into SU(3)
octets and decouplets had immense success. The use of first order
perturbation in explicit symmetry breaking, surprising at first
sight because of large mass differences, led to dramatic
predictions such as the existence  of the pseudoscalar meson
$\eta$ and the strange baryon $\Omega$. Their subsequent
experimental verification combined with Cabibbo's explanation of
the weak decays in terms of a mixing angle~\cite{cabbibo}, and numerous other
successes led eventually to the quark model and quantum
chromodynamics as the theory of strong interactions.

The subsequent discoveries of baryons and mesons having flavors other
than strangeness  (charm, beauty and top) have raised an
interesting, perhaps hypothetical question: to what extent the
success of broken SU(3) flavor symmetry stretches beyond the
strange quark mass? With this in mind, we investigate the
predictions of SU(3) flavor symmetries with $beauty$ and
$charm$ instead of $strangeness$ as the flavor. We use the
Gell-Mann-Okubo mass formulas~\cite{gmo} to relate the masses and predict new states to be discovered.

The generic mass formulas in the case of strange baryons and
mesons based on an explicit symmetry breaking term that transforms
like an octet are as follows:

Baryons:
\begin{equation}
3M_\Lambda  + M_\Sigma= 2 (M_N +M_\Xi)
\label{gmo}
\end{equation}

Mesons(pseudoscalar):
\begin{equation}
3 m_{\eta_8}^2+m_\pi^2  = 4 m_K^2  
\label{gmo2}
\end{equation}

Mesons (Vector)
\begin{equation}
3 m_{\omega_8}^2 +m_\rho^2 =  4 m_{K*}^2
\label{gmov}
\end{equation}

The spin 3/2 decouplet members ($ N^*, \Sigma^*, \Xi^*, \Omega$)
obey an equal spacing rule.

Also, in our analysis, we need the standard single-octet mixing
angle. In the case of the generic pseudoscalar octet, the singlet
and octet states are governed by the equations;
\begin{eqnarray}
|\eta\rangle = |\eta_8\rangle cos\theta  + |\eta_1\rangle sin\theta 
\nonumber \\
|\eta^\prime \rangle = -|\eta_8\rangle sin\theta + |\eta_1\rangle cos\theta
\label{mix}
\end{eqnarray}

On the other hand, the physical masses are related by the equations,

\begin{eqnarray}
M_{\eta_8} = M_{\eta}cos^2(\theta)+M_{\eta'}sin^2(\theta), 
\nonumber \\
M_{\eta_1} = M_{\eta}sin^2(\theta)+M_{\eta'}cos^2(\theta)
\label{mass}
\end{eqnarray}

Recall how these mass breaking and mixing {\it ans\"{a}tze} are applied in the ordinary octet/singlet pseudoscalar mesons. The GMO formula for mesons, Eq.~\ref{gmo2}, gives M($\eta_8$)= 566.7 GeV. Then Eq.~\ref{mass} can be solved via
\begin{equation}
cos^2(\theta)=\frac{M(\eta_8) -Ð M(\eta^\prime)}{M(\eta) -Ð M(\eta^\prime)}  
\label{cos}
\end{equation}
to yield $\theta$=-12.5$^\circ$ and M($\eta_1$) = 938.6 GeV . There are other approaches to obtaining the mixing angle, particularly through comparison of $\eta$ and $\pi^0$ into $\gamma \gamma$ decays. We will keep the value here, for consistency with the following applications of mixing and symmetry breaking. The sign of the angle is usually chosen as negative to produce certain decay interference effects, although here it is not relevant.
\vspace{1cm}

\noindent {\bf Mesons with $beauty$}

\begin{figure}
\vspace{1.7in}
\includegraphics{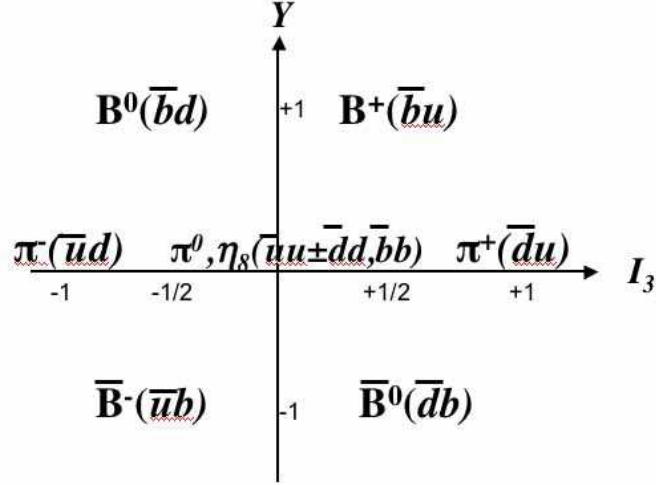}
\vspace{0.3in}
\caption{Pseudoscalar meson octet representation of $SU(3)_B$.}
\label{plot1}
\end{figure}

We now posit a new symmetry, SU(3)$_B$, in which the conventional assignments of flavor are altered by replacing the {\it s-flavor} by {\it b-flavor}. Then for the meson representations as shown in Fig.~\ref{plot1},
$B^{\pm}$, $B^0$ and $\bar{B}^0$ are the natural counterparts to the 
$K^{\pm}$, $K^0$ and $\bar{K}^0$, forming iso-doublet components of an octet along with $\pi^{\pm}$ and $\eta_B, \eta_B^\prime$. We apply this to the pseudoscalars. 

From Particle Data Group listings~\cite{pdg} the masses are
\begin{eqnarray}
M(B^{\pm}) = 5279\pm0.5 \, \rm{MeV} \nonumber \\
M(B^0) = 5279\pm0.5 \, \rm{MeV} \nonumber \\
M(\eta_B^\prime) = 9300\pm20 \, \rm{MeV} 
\label{massb}
\end{eqnarray}

Note that we choose the unconfirmed 9300 state to be the $\eta_B^\prime$. This is in the vicinity of the expected state, although it has not been settled upon experimentally~\cite{bbbar}. We would expect a lower mass state to contain the $u$ and $d$ flavors in this octet. Using these values and the GMO formula of Eq.~\ref{gmo2}, we obtain 

$$M(\eta_{B8})= 6095.1 \, \rm{MeV}.$$

Now from the masses of $\eta_{B8}$ and our assumed $\eta_B^\prime$, along with the pseudoscalar mixing angle we determine the lower mass physical $\eta_B$,

$$M(\eta_B)=5938 \, \rm{MeV} \,\,\,
\rm{and} \,\, M(\eta_{B1}) = 9142 \, \rm{MeV}.$$

Note that if we let the $\eta_B^\prime$ mass go to 9400, just 60 MeV below the $\Upsilon$, the $\eta_B$ drops by only 5 MeV. The $\eta_B$ predicted should be observable; a neutral $J^{PC}=0^{-+}$ with significant hidden $beauty$ at a mass well below the $\Upsilon$. Given the admixture of hidden $u$ and $d$ flavors there are many open decay channels, so decay width will be very broad. Its production via $\Upsilon\rightarrow\gamma+\eta_B$ should be quite striking.

We will not apply the symmetry to the vector mesons. There the mixing angle for the usual flavor SU(3) octet is quite close to the ideal value, for which the $\phi$ is purely a hidden strange state ($\bar{s}s$), satisfying the Zweig rule for its decays. This is true for the ground state charmonium and bottomonium vector mesons as well.
\vspace{0.5cm}

{\bf Baryons with {\it beauty}:}

The following baryonic states have been experimentally
established: 

\noindent Spin $\frac{1}{2}$ b-Baryons:

        $\Lambda_b^0$(5620 MeV); $\Sigma_b^-$(5816 MeV);  $\Sigma_b^+$(5808 MeV) \\

\noindent Spin $\frac{3}{2}$  b-Baryons:

        $\Sigma_b*$(5829 MeV);      $\Sigma_b^0 - \Delta$ = 4597 MeV.\\

\begin{figure}
\vspace{1.7in}
\includegraphics{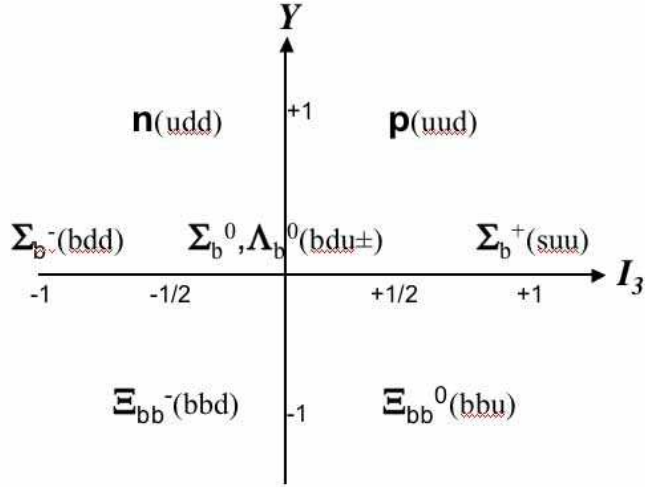}
\vspace{0.3in}
\caption{Spin $\frac{1}{2}$ baryon octet representation of $SU(3)_B$.}
\label{plot2}
\end{figure}

Again we replace {\it s-flavor} with {\it b-flavor} to form SU(3)$_B$ as shown in Fig.~\ref{plot2}. Using the GMO linear formula Eq.~\ref{gmo} ( with N as the member of the Octet), and
using the values for $\Lambda_b^0$(5620) and the average of
$\Sigma_b^-$(5816 MeV);  $\Sigma_b^+$(5808 MeV) we find,
$M_{\Xi_{bb}}^{0,-}$ = 10,400 MeV.
\vspace{1cm}

\begin{figure}
\vspace{1.7in}
\includegraphics{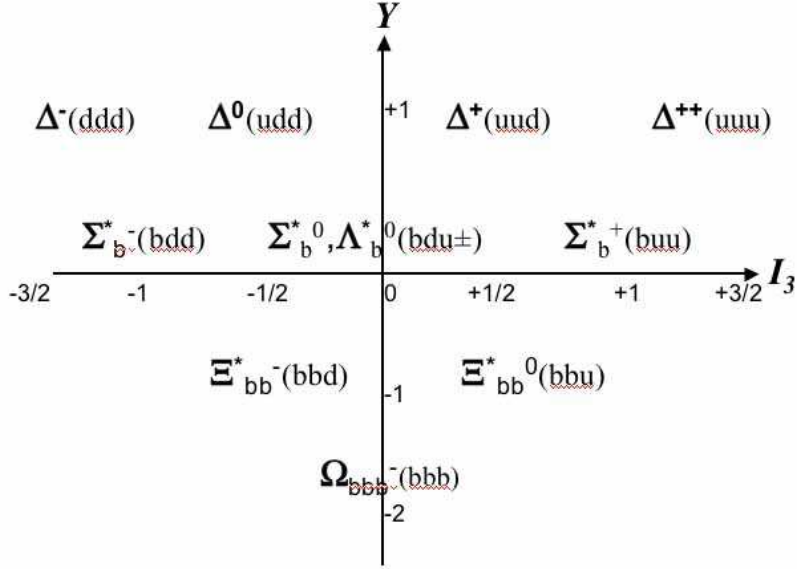}
\vspace{0.3in}
\caption{Spin $\frac{3}{2}$ baryon decuplet representation of $SU(3)_B$.}
\label{plot3}
\end{figure}

Using the decuplet equal spacing rule (N$^*$ or $\Delta$ as the member) for the representation in Fig.~\ref{plot3}, we
predict the masses of the other two members, $M(\Xi_{bb}^*)$=10,426
MeV and $M(\Omega_{bbb})$=15,023 MeV. These predictions await experimental discovery of these multiple beauty states. 

\vspace{1cm}

{\bf {\it Charm} and SU(3) multiplets}

The above SU(3)$_B$ multiplets were obtained by substituting the $b-$flavor for the $s-$flavor. What about the $c-$flavor, with associated charge +2/3? It makes some sense to replace $u-$flavor by $charm$ to form SU(3)$_C$. Then, however, the normal octet assignments for the mesons would involve large mass breaking for equal hypercharge Y states, {\it i.e.} fixed ``I-spin" states. On the other hand, we know that there is smaller splitting among equal charm states. This suggests that the octet for (c,d,s) flavors be a ``U-spin" octet. The states are assigned with charge on the vertical axis and $U_3$ on the horizontal axis, as indicated by flavor labels in Fig.~\ref{plot4} for pseudoscalars.

The masses of these pseudoscalars are:
\begin{figure}
\vspace{2.0in}
\includegraphics{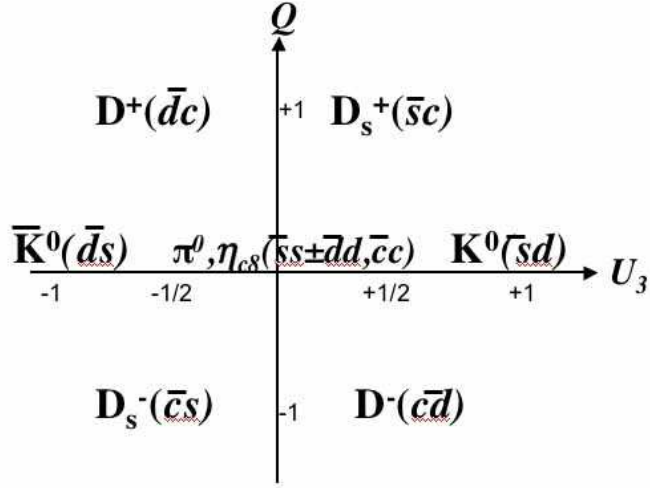}
\vspace{0.3in}
\caption{U-spin pseudoscalar meson octet representation of $SU(3)_c$.}
\label{plot4}
\end{figure}

\begin{eqnarray}
M(D^+[\bar{d}c])=1869 \rm{MeV}\,\,\,\,\,\,M(D_s^+[\bar{s}c])=1968 \rm{MeV} 
\nonumber \\
M(\bar{K}^0[\bar{d}s])=498 \rm{MeV}=M(K^0[\bar{s}d]).
\label{uspin}
\end{eqnarray}
Taking the average for $D^+$ and $D_s^+$ and applying the GMO formula Eq.~\ref{gmo2}, we obtain an $\eta_{c8}$ mass of 2197.1 MeV. What about the physical $\eta$ states? There is a signal at 2100 MeV~\cite{pdg} that we can associate with the lower mass state $\eta_c$ (not to be confused with the state at 2980 MeV). Then with the same pseudoscalar mixing angle of 12.5$^\circ$, we obtain high mass states

$$M(\eta_c^\prime)=4171\,\rm{MeV} \,\,\rm{and}\,\,M(\eta_{c1})=4074\,\rm{MeV}.$$  

It is worth noting that for a somewhat larger mixing angle of 19.4$^\circ$ the $\eta_c^\prime$ would drop down to 2980 MeV, where the known charmonium pseudoscalar lies. But the anchor here remains the tentative $\eta_c$ at 2100 MeV. If this state is established to be a {\it bone fide} 0$^{-+}$, it will be a state with significant hidden charm. 

\begin{figure}
\vspace{1.7in}
\includegraphics{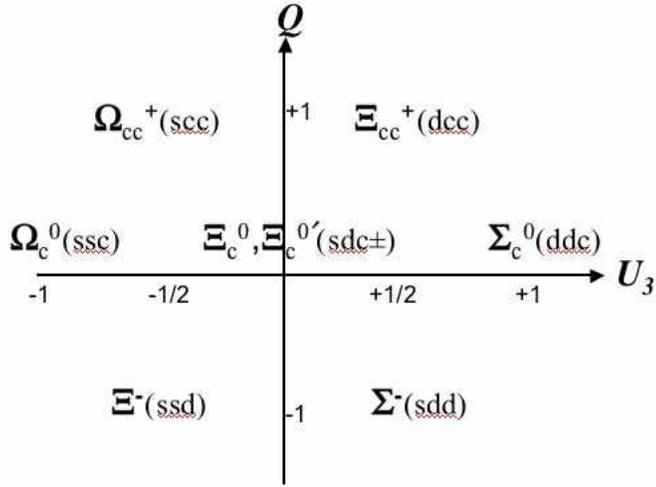}
\vspace{0.3in}
\caption{U-spin and Spin $\frac{1}{2}$ baryon octet representation of $SU(3)_b$.}
\label{plot5}
\end{figure}

Charmed baryons in this SU(3)$_C$ scheme fall into octets and anti-decuplets. 
We will not deal with the decuplet, since there are very few established $\frac{3}{2}^+$ charmed baryons. For the octet of Fig.~\ref{plot5}, we can use the mass breaking as for the mesons. The U-spin multiplets are preserved, although within each U-spin multiplet there is some breaking, albeit smaller than the breaking from one charm level to the next. The $\Omega_{cc}(s c c)^+$ has not been reported, but there is a signal for $\Xi_{cc}(d c c)^+$ at 3519 GeV. This can be considered as the ``anchor" state for the mass breaking. Using this mass along with $M(\Sigma_c^0)=2454$ GeV, $M(\Omega_c^0)=2698$ GeV and two states $M(\Xi_c^0)=2471$ GeV and $2578$ GeV, we evaluate the GMO formula to obtain the $\Omega_{cc}^+$ mass at $3953$ GeV. This puts the charm 2 state within reach, experimentally.

What about top-flavor states? Because of the high mass of the top quark there is a very rapid decay into beauty states and other hadrons, before top hadron states, baryons or mesons, can form. 

Finally, in the list of tentative meson and baryon states~\cite{pdg} there are other possible candidates for our ``anchor" states. In conventional quark model assignments these intermediate mass states fall through the cracks. They may very well be evidence of residual SU(3) symmetries that codify mass breaking more readily than the conventional view. This would not invalidate the conventional non-relativistic quark models, based on QCD expectations, and their extensive agreements with data. Nor would the existence of intermediate mass states demonstrate the correctness of the SU(3)$_{flavor}$ that we considered here. However the SU(3)$_{B\, \rm{or}\, C}$ would provide a useful phenomenological model of mass breaking among the heavy flavor hadrons.

\end{document}